\documentclass[aps,prb,twocolumn]{revtex4}
\usepackage{amsmath,amssymb,mathrsfs}

\begin{document}

\title{Theoretical and numerical investigation of the size-dependent
  optical effects in metal nanoparticles}

\author{Alexander A. Govyadinov~\footnote{Formerly, at the
    Department of Bioengineering, University of Pennsylvania,
    Philadelphia, PA 19104}}

\affiliation{CIC nanoGUNE Consolider, Avenida de Tolosa 76, Guipuzcoa 20018, Spain}

\author{George Y. Panasyuk~$^{*}$}

\affiliation{Propulsion Directorate, Air Force Research Laboratory,
  Wright-Patterson Air Force Base, OH 45433} 

\author{John C. Schotland}

\affiliation{Department of Mathematics, University of Michigan, Ann Arbor, MI 48109} 

\author{Vadim A. Markel} 

\affiliation{Departments of Radiology and Bioengineering and the
  Graduate Group in Applied Mathematics and Computational Science,
  University of Pennsylvania, Philadelphia, Pennsylvania 19104}

\begin{abstract}
  We further develop the theory of quantum finite-size effects in
  metallic nanoparticles, which was originally formulated by Hache,
  Ricard and Flytzanis [J. Opt. Soc. Am. B {\bf 3}, 1647 (1986)] and
  (in a somewhat corrected form) by Rautian [Sov. Phys. JETP {\bf 85},
  451 (1997)]. These references consider a metal nanoparticle as a
  degenerate Fermi gas of conduction electrons in an infinitely-high
  spherical potential well. This model (referred to as the HRFR model
  below) yields mathematical expressions for the linear and the
  third-order nonlinear polarizabilities of a nanoparticle in terms of
  infinite nested series. These series have not been evaluated
  numerically so far and, in the case of nonlinear polarizability,
  they can not be evaluated with the use of conventional computers due
  to the high computational complexity involved.  Rautian has derived
  a set of remarkable analytical approximations to the series but
  direct numerical verification of Rautian's approximate formulas
  remained a formidable challenge. In this work, we derive an
  expression for the third-order nonlinear polarizability, which is
  exact within the HRFR model but amenable to numerical
  implementation. We then evaluate the expressions obtained by us
  numerically for both linear and nonlinear polarizabilities. We
  investigate the limits of applicability of Rautian's approximations
  and find that they are surprizingly accurate in a wide range of
  physical parameters. We also discuss the limits of small frequencies
  (comparable or below the Drude relaxation constant) and of large
  particle sizes (the bulk limit) and show that these limits are
  problematic for the HRFR model, irrespectively of any additional
  approximations used. Finally, we compare the HRFR model to the
  purely classical theory of nonlinear polarization of metal
  nanoparticles developed by us earlier [Phys.  Rev. Lett. {\bf 100},
  47402 (2008)].
\end{abstract}

\date{\today}

\maketitle 

\section{Introduction}
\label{sec:intro}

This paper is dedicated to the memory of Sergey Glebovich Rautian
(1928-2009) who was a teacher to some of us and inspiration to all.

Metal nanoparticles have received an extraordinary amount of attention
recently because of their ability to greatly enhance local fields.
The enhancement is attributed to the excitation of surface plasmons
and it has a variety of applications in
photovoltaics~\cite{photovoltaics}, sensing~\cite{chemSensing} and
surface-enhanced Raman
scattering~\cite{SERSkelly,SERSelSayed,SERShao}. Currently,
nanopartices of very small sizes, up to a few nanometers, are
customarily used in experiments. The theoretical description of the
optical properties of such nanoparticles is most frequently based on
the macroscopic electrodynamics. At least, this is typical in the
field of plasmonics. However, macroscopic electrodynamics can not
capture certain effects of finite size. Hache, Ricard and
Flytzanis~\cite{hache_86_1} and Rautian (in a somewhat modified
form)~\cite{rautian_97_1} have developed an elaborate theory of
quantum finite-size effects in metal nanoparticles. In
Refs.~\onlinecite{hache_86_1,rautian_97_1}, a nanoparticle was modeled
as degenerate Fermi gas confined in an infinite potential well of
spherical shape (below, the HRFR model).  Despite being fairly simple,
the HRFR model results in very complicated formulas, which can not be
evaluated numerically even with the aid of modern computers. For
example, the expression for the third-order nonlinear polarizability
involves a twelve-fold nested summation. Rautian has reduced the
number of summations from twelve to eight by performing summation over
the magnetic sublevels analytically; he then obtained a number of
remarkable approximations to the resulting eight-fold 
summation~\cite{rautian_97_1}. However, these approximations have
never been verified directly due to the overwhelming numerical
complexity involved.

In this contribution, we develop the analytical theory of Rautian a
step further by reducing the number of nested summations involved from
eight to five without making any additional approximations. This turns
out to be sufficient to render the formulas amenable to direct
numerical implementation. We then compare the results of numerical
evaluation of the five-fold series derived by us to the results, which
follow from Rautian's approximate formulas, and discuss various
physical limits, including the limits of low frequency and large
particle size.

In Sec.~\ref{sec:Rautian} we review Rautian's theory. Here we use
somewhat simplified notation and, in particular, avoid the use of
irreducible spherical tensors and $6j$-symbols. In
Sec.~\ref{sec:further}, we develop the theory further by utilizing the
orbital selection rules and reduce the nested summation involved in
the definition of the third-order nonlinear polarizability from
eight-fold to five-fold. In Sec.~\ref{sec:Ei_Ee}, we describe a simple
method to relate the internal and applied fields, which is to the
first order consistent with the approach proposed in
Ref.~\onlinecite{drachev_04_2}, but is more mathematically rigorous.
In Sec.~\ref{sec:num}, the results of numerical computations are
reported.  A summary of obtained results and a discussion are
contained in Sec.~\ref{sec:summ}.

\section{Rautian's theory}
\label{sec:Rautian}

We start by reviewing Rautian's theory of quantum finite-size effects
in conducting nanoparticles~\cite{rautian_97_1}. The physical system
under consideration is a gas of ${\mathscr N}$ non-interacting
electrons placed inside a spherical, infinitely deep potential well of
radius $a$ and subjected to harmonically-oscillating,
spatially-uniform electric field
\begin{equation}
\label{Ei_def}
{\bf E}_i(t) = {\bf A}_i \exp(-i \omega t) + {\rm c.c.}
\end{equation}
\noindent
Note that ${\bf E}_i$ is the electric field {\em inside} the
nanoparticle. It will be related to the {\em external} (applied) field
${\bf E}_e$ in Sec.~\ref{sec:Ei_Ee} below.

Since the nanoparticle is assumed to be electrically
small (that is, $a \ll \lambda = 2\pi c /\omega$), the electron-field
interaction can be described in the dipole approximation by the
time-dependent operator
\begin{equation}
\label{eq:V}
V(t) = - {\bf d} \cdot {\bf E}(t) \ ,
\end{equation}
\noindent
where ${\bf d}$ is the dipole moment operator. Under the additional
assumption that ${\bf E}_i(t)$ is linearly polarized with a purely
real amplitude ${\bf A}_i = A_i \hat{\bf z}$, we can write
\begin{equation}
\label{eq:Vlin}
V(t) =  G \exp(-i\omega t) + {\rm c.c.} \ ,
\end{equation}
\noindent
where
\begin{equation}
\label{G_def}
G = - {\bf d} \cdot {\bf A}_i \ .
\end{equation}
Rautian made use of the interaction representation in which the wave
function is expanded in the basis of the unperturbed Hamiltonian
eigenstates. The single-electron unperturbed states are
\begin{equation}
\label{phi_def}
\phi_{nlm}({\bf r}) = \frac{1}{Z_{nl}}j_l\left(\xi_{nl} r/a \right) Y_{lm}(\hat{\bf r}) \ ,
\end{equation}
\noindent
where $j_l(x)$ are the spherical Bessel functions of the first kind
and order $l$; $\xi_{nl}$ is the $n$-th positive root ($n=1,2,\ldots$)
of the equation $j_l(x)=0$, $Y_{lm}(\hat{\bf r})$ are the spherical
functions (viewed here as functions of the polar and azimuthal angles
of the unit vector $\hat{\bf r} = {\bf r} / r$) and
\begin{equation}
\label{Z_def}
Z_{nl} = \sqrt{\frac{a^3}{2}}j_{l+1}(\xi_{nl})
\end{equation}
\noindent
are normalization factors. The energy eigenstates are labeled by the
main quantum number $n$, the orbital number $l$ and the magnetic
number $m$. The unperturbed energy levels are given by the formula
\begin{equation}
E_{nl} = E_0 \xi_{nl}^2 \ ,
\end{equation}
\noindent
where 
\begin{equation}
E_0 = \frac{\hbar^2}{2 m_e a^2}
\end{equation}
\noindent
and $m_e$ is the electron mass.

In what follows, we use the composite indices $\mu$, $\nu$, $\eta$ and
$\zeta$ to label the eigenstates. Each composite index corresponds to
the triplet of quantum numbers $(n,l,m)$. By convention, if
$\mu=(n,l,m)$, then $\mu^\prime = (n^\prime,l^\prime,m^\prime)$. The
matrix elements of the $z$-projection of the dipole moment operator
are given by
\begin{equation}
\label{eq:d_del}
\hat{\bf z} \cdot {\bf d}_{\mu \mu^\prime} = ea \Delta_{\mu
  \mu^\prime} \ ,
\end{equation}
\noindent
where 
\begin{equation}
\label{eq:del}
\Delta_{\mu \mu^\prime} = \delta_{m m^\prime} R_{nl}^{n^\prime
  l^\prime} \left( b_{lm} \delta_{l-1,l^\prime} + b_{l+1,m} \delta_{l+1,l^\prime} \right) \
, 
\end{equation}
$\delta_{ll^\prime}$ are the Kronecker delta-symbols, $l,l^\prime\geq
0$ and
\begin{equation}
\label{b_R_def}
b_{lm}  = \sqrt{\frac{l^2-m^2}{4l^2-1}} \ , \ \
R_{nl}^{n^\prime l^\prime} = \frac{4 \xi_{nl} \xi_{n^\prime l^\prime}}
{\left( \xi_{nl}^2 - \xi_{n^\prime l^\prime}^2 \right)^2} \ .
\end{equation}
\noindent
Note that the diagonal elements of $\Delta$ are all equal to zero, as
is the case for any system with a center of symmetry. Finally, the
matrix elements of the operator $G$ are given by
\begin{equation}
\label{G_A}
G_{\mu\mu^\prime} = - ea \Delta_{\mu\mu^\prime} A_i \ .
\end{equation}

The density matrix of the system, $\rho$, can be written in the form
$\rho_{\mu\nu}(t) = \tilde{\rho}_{\mu\nu}(t) \exp(i\omega_{\mu\nu}t)$,
where $\omega_{\mu\nu} = (E_\mu - E_\nu)/\hbar$ are the transition
frequencies and $\tilde{\rho}_{\mu\nu}(t)$ is the so-called
slow-varying amplitude, which obeys the following master
equation~\cite{rautian_book_91}:
\begin{align}
\label{eq:rhoDifEq}
\left(\frac{\partial}{\partial t} + i \omega_{\mu\nu} +
  \Gamma_{\mu\nu}\right) \tilde{\rho}_{\mu\nu} =
 \delta_{\mu\nu} \Gamma_{\mu\mu} N_\mu \nonumber \\
- \frac{i}{\hbar}\sum_\eta \left[ V_{\mu\eta}(t)
  \tilde{\rho}_{\eta\nu} - \tilde{\rho}_{\mu\eta} V_{\eta\nu}(t)
\right] \ .
\end{align}
\noindent
Here $N_\mu$ are the equilibrium state populations and
$\Gamma_{\mu\nu}$ are phenomenological relaxation constants. Following
Rautian, we assume that
\begin{equation}
\label{Gamma_1_2}
\Gamma_{\mu\nu} = \Gamma_1 \delta_{\mu\nu} + \Gamma_2 (1 -
\delta_{\mu\nu}) \ . 
\end{equation}
\noindent
Eq.~(\ref{Gamma_1_2}) is the least complex assumption on
$\Gamma_{\mu\nu}$, which still distinguishes the relaxation rates for
the diagonal and off-diagonal elements of the density matrix.

It can be seen that, for the case of zero external field,
$\tilde{\rho}_{\mu\nu} = \delta_{\mu\nu} N_\mu$. The Fermi statistics
is introduced at this point by writing
\begin{equation}
\label{Fermi}
N_\mu= \frac{2}{\exp[(E_\mu - E_F)/(k_B T)] + 1} \ ,
\end{equation}
\noindent
where $E_F$ is the Fermi energy, $k_B$ is Boltzmann's constant, $T$ is
the temperature, and the factor of two in the numerator accounts for
the electron spin. Conservation of particles reads $\sum_\mu N_\mu =
{\mathscr N}$. When ${\mathscr N}\gg 1$, the well-known analytical
formula for the Fermi's energy,
\begin{equation}
\label{E_F}
E_F = E_0 \left( 3\pi^2 \right)^{2/3} \left( \frac{a}{\ell} \right)^2
= \left( 3\pi^2 \right)^{2/3} \frac{\hbar^2}{2 m_e \ell^2} \ ,
\end{equation}
\noindent
holds with a good accuracy. Here $\ell$ is the characteristic atomic
scale, defined by the relation
\begin{equation}
\label{ell_def}
\ell^3 = \Omega / {\mathscr N} \ ,
\end{equation}
where
\begin{equation}
\label{Omega_def}
\Omega = 4\pi a^3 / 3
\end{equation}
\noindent
is the nanoparticle volume. Thus, $\ell^3$ is the specific volume per
conduction electron. We note that $\ell$ is, generally, different from
the lattice constant $h$. Many metals of interest in plasmonics have
an fcc lattice structure with four conduction electrons per unit cell.
In this case $h = 4^{1/3}\ell$. For example, in silver, $\ell \approx
0.26{\rm nm}$, $h\approx 0.41{\rm nm}$ and $E_F\approx 5.51{\rm eV}$.
At room temperature ($T=300{\rm K}$), $k_B T \approx 0.026{\rm eV}$,
so that $T=0$ is a good approximation. In this case, $N_\mu = 2$ if
$E_\mu\leq E_F$ and $N_\mu = 0$ otherwise.  Most numerical results
shown below have been obtained in this limit.  However, to illustrate
the effects of finite temperature, we have also performed some
computations at $T=300{\rm K}$. Finally, the Fermi velocity is given
by the equation
\begin{equation}
\label{v_F_def}
v_F = \sqrt{\frac{2E_F}{m_e}} = (3\pi^2)^{1/3} \frac{\hbar}{m_e \ell} \ .
\end{equation}
\noindent
In silver, $v_F\approx 1.2\cdot 10^8 {\rm cm/sec}$ and,
correspondingly, $c/v_F \approx 250$. Electron velocities in excited
states are expected to be no larger than a few times Fermi velocity,
still much smaller than $c$. This justifies the use of
non-relativistic quantum mechanics.

The solution to (\ref{eq:rhoDifEq}) has the form of a Fourier series:
\begin{equation}
\label{eq:rhoSer}
\tilde{\rho}_{\mu\nu}(t) = \sum_{s=-\infty}^\infty \tilde{\rho}_{\mu\nu}^{(s)}\exp(-i s\omega t) \ .
\end{equation}
\noindent
The expansion coefficients $\tilde{\rho}_{\mu\nu}^{(s)}$ obey the system of
equations:
\begin{align}
\label{eq:rhos}
\tilde{\rho}_{\mu\nu}^{(s)}  = \delta_{\mu\nu} \delta_{s0} N_{\mu} & -
\frac{\Lambda_{\mu\nu}^{(s)}(\omega)}{\hbar\omega}  \sum_\eta \left[
  G_{\mu\eta} \left( \tilde{\rho}_{\eta\nu}^{(s-1)} +
    \tilde{\rho}_{\eta\nu}^{(s+1)}\right) \right. \nonumber \\
& - \left. G_{\eta\nu} \left( \tilde{\rho}_{\mu\eta}^{(s
      - 1)} + \tilde{\rho}_{\mu\eta}^{(s+1)} \right) \right] \ ,
\end{align}
\noindent
where
\begin{equation}
\label{Lambda_def}
\Lambda_{\mu\nu}^{(s)}(\omega) =
\frac{\omega}{\omega_{\mu\nu} - s\omega - i \Gamma_{\mu\nu}}
\end{equation}
\noindent
are Lorentzian spectral factors.

The optical response of the nanoparticle is determined by the
quantum-mechanical expectation of its total dipole moment, which is
given by
\begin{equation}
\label{eq:dExpect}
\langle d(t) \rangle = ea \sum_{\mu\nu} \Delta_{\mu\nu}
\tilde{\rho}_{\mu\nu} (t) \ . 
\end{equation}
\noindent
Upon substitution of (\ref{eq:rhoSer}) into (\ref{eq:dExpect}), we
obtain the expansion of $\langle d(t) \rangle$ into temporal Fourier
harmonics. We now consider the optical response at the fundamental
frequency $\omega$, which describes degenerate nonlinear phenomena
such as the four-wave mixing. Denoting the component of $\langle d(t)
\rangle$, which oscillates at the frequency $\omega$, by $\langle
d_\omega(t) \rangle$, we can write
\begin{equation}
\label{eq:dwt}
\langle d_\omega (t) \rangle = D \exp(-i\omega t) + {\rm c.c.} \ ,
\end{equation}
\noindent
where 
\begin{equation}
\label{D_def}
D = ea \sum_{\mu\nu} \Delta_{\mu\nu}\tilde{\rho}_{\mu\nu}^{(1)} \ .
\end{equation}
\noindent
The coefficients $\tilde{\rho}_{\mu\nu}^{(1)}$ and the amplitude $D$
in (\ref{eq:dwt}) can be expanded in powers of $A_i$. Namely, we can
write
\begin{equation}
\label{chi_nk}
D = \Omega \chi(A_i) A_i \ ,
\end{equation}
\noindent
where
\begin{equation}
\label{eq:D}
\chi(A_i) = \chi_1 + \chi_3 (A_i/A_{\rm at})^2 + \chi_5 (A_i/A_{\rm at})^4 \ldots 
\end{equation}
\noindent
Here we have introduced the characteristic atomic field
\begin{equation}
\label{A_at_def}
A_{\rm at} = e/\ell^2
\end{equation}
\noindent
and have used the assumption that $A_i$ is real-valued; in the more
general case, the expansion contains the terms of the form $\chi_3
\vert A_i/A_{\rm at} \vert^2$, etc. Note that the definition of
$\chi_3$ in (\ref{chi_nk}),(\ref{eq:D}) is somewhat unconventional.
The nonlinear susceptibility $\chi^{(3)}$, as defined in standard
expositions of the subject~\cite{boyd_book_92}, has the dimensionality
of the inverse square of the electric field, that is, of ${\rm
  cm}^2/{\rm statvolt}^2 = {\rm cm}^3/{\rm erg}$ in the Gaussian
system of units. Here we find it more expedient to define
dimensionless coefficients $\chi_1$, $\chi_3$, $\chi_5$, etc., and
expand the dipole moment amplitude $D$ in powers of the dimensionless
variable $A_i/A_{\rm at}$.

The first two coefficients in the expansion (\ref{eq:D}), $\chi_1$ and
$\chi_3$, have been computed by Rautian explicitly and are given by
the following series:
\begin{subequations}
\label{chi_1_3}
\begin{align}
\label{eq:chi1}
\chi_1 &= \frac{(ea)^2}{(\hbar\omega) \Omega} \sum_{\mu\nu} N_{\mu\nu}
\Lambda_{\mu\nu}^{(1)} \Delta_{\mu\nu} \Delta_{\nu\mu} \ , \\ 
\label{eq:chi3}
\chi_3 &= \frac{(ea)^4 A_{\rm at}^2}{(\hbar\omega)^3\Omega}
\sum_{\mu\nu\eta\zeta} B_{\mu\nu}^{\zeta\eta} \Delta_{\mu\zeta}
\Delta_{\zeta\eta} \Delta_{\eta\nu} \Delta_{\nu\mu} \ ,
\end{align}
\end{subequations}
\noindent
where
\begin{align}
  B_{\mu\nu}^{\zeta\eta} & = \Lambda_{\mu\nu}^{(1)} N_{\zeta\mu}
  \left[ {\Lambda_{\mu\eta}^{(0)} \left(\Lambda_{\mu\zeta}^{(1)} +
        \Lambda_{\mu\zeta}^{(-1)}\right) + \Lambda_{\mu\eta}^{(2)}
      \Lambda_{\mu\zeta}^{(1)} } \right]
  \nonumber \\
  & + \Lambda_{\mu\nu}^{(1)} N_{\nu\eta} \left[
    \Lambda_{\zeta\nu}^{(0)} \left( \Lambda_{\eta\nu}^{(1)} +
      \Lambda_{\eta\nu}^{(-1)} \right) + \Lambda_{\zeta\nu}^{(2)}
    \Lambda_{\eta\nu}^{(1)} \right] \nonumber \\
  & - \Lambda_{\mu\nu}^{(1)} N_{\eta\zeta} \left(
    \Lambda_{\zeta\eta}^{(1)} + \Lambda_{\zeta\eta}^{(-1)} \right)
  \left( \Lambda_{\zeta\nu}^{(0)} + \Lambda_{\mu\eta}^{(0)} \right)
  \nonumber \\
  & - \Lambda_{\mu\nu}^{(1)} N_{\eta\zeta} \Lambda_{\zeta\eta}^{(1)}
  \left( \Lambda_{\zeta\nu}^{(2)} + \Lambda_{\mu\eta}^{(2)} \right) \ .
\label{eq:B}
\end{align}

\noindent
Here $N_{\mu\nu}=N_\mu-N_\nu$. Note that all quantities inside the
summation symbols are dimensionless and so are the factors in front of
the summation symbols.

The expression (\ref{eq:chi3}) involves a staggering 12-fold summation
(recall that each composite index $\mu$, $\nu$, $\eta$ and $\zeta$
consists of three integer indices). Rautian used the mathematical
formalism of irreducible spherical tensors and $6j$-symbols to perform
summation over magnetic sublevels analytically and to reduce the
expression to an 8-fold summation. However, this approach does not
make use of the orbital selection rules, which are explicit in
(\ref{eq:del}). In Sec.~\ref{sec:further}, we will use the orbital
selection rules to analytically reduce (\ref{eq:chi3}) to a 5-fold
summation. The resultant formula is amenable to direct numerical
implementation, as will be illustrated in Sec.~\ref{sec:num}.

Having performed the summation over the magnetic sublevels, Rautian
has evaluated the resulting series by exploiting the following two
approximations:

\begin{enumerate}
\item Adopt the two-level approximation~\cite{hache_86_1}. In this
  approximation, only the terms with $\mu=\eta$ and $\nu=\zeta$ are
  retained in the right-hand side of (\ref{eq:chi3}).
\item Assume that there are two dominant contributions to the series
  (\ref{chi_1_3}). The off-resonant (Drudean) contribution is obtained
  by keeping only the terms with $\omega_{\mu\nu} \ll \omega$ in the
  Lorentzian factors $\Lambda_{\mu\nu}^{(s)}(\omega)$. The resonant
  contribution is obtained by keeping only the terms with
  $\omega_{\mu\nu} \approx \omega$. Each contribution is then
  evaluated separately by replacing summation with integration.
\end{enumerate}

Using the same approximations, we have reproduced Rautian's analytical
results. For $\chi_1$, we obtained
\begin{equation}
\label{chi_1_an}
\chi_1 =  -\frac{1}{4\pi} \frac{\omega_p^2}{\omega + i \Gamma_2}
\left[ \frac{F_1}{\omega + i \Gamma_2} - i g_1 \frac{v_F/a}{\omega^2} \right] \ ,
\end{equation}
\noindent
where 
\begin{equation}
\label{omega_p_def}
\omega_p = \sqrt{ \frac{4\pi e^2}{m_e \ell^3}} 
\end{equation}
\noindent
is the plasma frequency and $F_1$, $g_1$ are dimensionless real-valued
functions, which weakly depend on the parameters of the problem and
are of the order of unity.  More specifically, $F_1$ is very close to
unity for all reasonable particle sizes and approaches unity
asymptotically when $a\rightarrow \infty$ (we have verified this
numerically). In what follows, we assume that $F_1=1$. The function
$g_1$ depends most profoundly on the ratio $\kappa=\hbar\omega/E_F$.
We can write, approximately,
\begin{equation}
\label{g1_def}
g_1 \approx \frac{1}{\kappa} \int_{1-\kappa}^1 x^{3/2}
(x+\kappa)^{1/2} dx \ , \ \ \kappa = \frac{\hbar \omega}{E_F} \ .
\end{equation}
An analytical expression for this integral and a plot are given in the
Appendix.

Equation (\ref{chi_1_an}) is equivalent to combining equations 3.16
and 3.23 of Ref.~\onlinecite{rautian_97_1}. On physical grounds, one can
argue that these expressions are applicable only if $\Gamma_2/\omega
\ll 1$. Indeed, in the classical Drude model, we have
\begin{equation}
\label{chi_1_D}
\chi_1^{\rm Drude} = -\frac{1}{4\pi} \frac{\omega_p^2}{\omega(\omega +
  i\gamma)} \ ,
\end{equation}
\noindent
where $\gamma$ is a relaxation constant. We expect the Drude model to
be accurate in the limit $a\rightarrow \infty$, when the second term
in the square brackets in (\ref{chi_1_an}) vanishes. Thus,
(\ref{chi_1_an}) has an incorrect low-frequency asymptote. We argue
that the asymptote is incorrect because the HRFR model disregards the
Hartree interaction potential. This will be discussed in more detail
in Sec.~\ref{sec:Ei_Ee} below. At this point, we assume that
$\Gamma_2/\omega \ll 1$ and expand (\ref{chi_1_an}) in
$\Gamma_2/\omega$, neglect the correction to the real part of the
resulting expression, and obtain:
\begin{equation}
\label{chi_1_exp}
\chi_1 \approx - \frac{1}{4\pi}\left(\frac{\omega_p}{\omega}\right)^2
\left[1 -  i \frac{2\Gamma_2 + g_1 v_F/a}{\omega} \right] \ .
\end{equation}
\noindent
This expression corresponds to equation 3.28 of
Ref.~\onlinecite{rautian_97_1}. Note that neglecting the correction to the
real part but retaining the correction to the imaginary part in the
above equation is mathematically justified because the terms
$2\Gamma_2$ and $v_F/a$ can be of the same order of magnitude, as we
will see below.

Comparing (\ref{chi_1_exp}) to a similar expansion of (\ref{chi_1_D}),
we conclude that the size-dependent relaxation constant $\gamma$ is
given by
\begin{equation}
\label{gamma_a}
\gamma \approx \gamma_\infty + g_1 \frac{v_F}{a} \ , \ \ \gamma_\infty = 2
\Gamma_2 \ ,
\end{equation}
\noindent
where $\gamma_\infty$ is the relaxation constant in bulk. It can be
seen that the ratio $v_F/a$ plays the role of the collision frequency.
The analytical result (\ref{gamma_a}) is widely known and frequently
used; it will be confirmed by direct numerical evaluation of
(\ref{eq:chi1}) below.

For the third-order nonlinear susceptibility $\chi_3$, we obtain, 
with the same accuracy as above, 
\begin{align}
\label{eq:Dan}
\chi_3 = & \frac{\Gamma_2}{\Gamma_1} \frac{\alpha^2}{10\pi^3} \left(
  \frac{a}{\ell} \right)^2 \left( \frac{\lambda_p}{\ell} \right)^2
\left( \frac{\omega_p}{\omega} \right)^4 \nonumber \\
\times & \left[ F_3 - i \left( F_3 \frac{\gamma_\infty}{\omega} + g_3
    \frac{(v_F/a)^5}{\omega^3 \gamma_\infty^2}\right) \right] \ .
\end{align}
\noindent
Here $\alpha = e^2/\hbar c \approx 1/137$ is the fine structure
constant, $\lambda_p =2\pi c/\omega_p$ is the wavelength at the plasma
frequency ($\approx 138{\rm nm}$ in silver) and $F_3$, $g_3$ is
another set of dimensionless real-valued functions of the order of
unity. For realistic parameters, the function $F_3$ varies only
slightly~\cite{rautian_97_1,drachev_04_2} between $0.30$ and $0.33$;
we have taken $F_3=0.33$ in the numerical computations of
Sec.~\ref{sec:num}. The function $g_3$ can be approximated by the
following integral:
\begin{equation}
\label{g3_def}
g_3 \approx \frac{1}{\kappa} \int_{1-\kappa}^1 x^{5/2}
(x+\kappa)^{3/2} dx \ , \ \ \kappa = \frac{\hbar \omega}{E_F} \ .
\end{equation}
The approximate formula (\ref{g3_def}) applies only for $\hbar\omega <
E_F$. However, we are interested in the spectral region $\omega
\lesssim \omega_p$. In silver, $\hbar\omega_p \approx 8.98{\rm eV}$
and $\hbar\omega_p/E_F\approx 1.63$. This leaves us with the spectral
range $E_F/\hbar < \omega < \omega_p$ in which (\ref{g3_def}) is not
applicable. The integral (\ref{g3_def}) can be evaluated analytically;
the resulting expression and plot are given in the Appendix.

Expression (\ref{eq:Dan}) contains several dimensionless parameters.
For a silver nanoparticle of the radius $a=10{\rm nm}$,
\begin{equation*}
\frac{\alpha^2}{10\pi^3} \left( \frac{a}{\ell} \right)^2 
                         \left( \frac{\lambda_p}{\ell} \right)^2
\approx 71.6 \ .
\end{equation*}
The ratio $\Gamma_2/\Gamma_1$ is more puzzling. While $\Gamma_2$ can
be related to the Drude relaxation constant through (\ref{gamma_a}),
$\Gamma_1$ does not enter the analytical approximations
(\ref{chi_1_an}), (\ref{chi_1_exp}) or the exact expression
(\ref{eq:chi1}). Therefore, $\Gamma_1$ can not be directly related to
any measurement of the linear optical response. It was previously
suggested~\cite{drachev_04_2} that, based on the available
experimental studies of non-equilibrium electron kinetics in
silver~\cite{groeneveld_95_1,delfatti_98_1,lehmann_00_1}, $\Gamma_2 /
\Gamma_1 \sim 10$. This ratio will be employed below.

Another interesting question is the dependence of the results on the
particle radius, $a$. It follows from the analytical approximation
(\ref{chi_1_exp}) that $\chi_1$ approaches a well-defined ``bulk''
limit when $a\rightarrow \infty$. The characteristic length scale is
$v_F/\gamma_\infty$ ($\approx 44{\rm nm}$ in silver). Of course,
direct numerical evaluation of $\chi_1$ according to (\ref{eq:chi1})
is expected to reveal some dependence of $\chi_1$ on $a$, which is not
contained in the analytical approximation (\ref{chi_1_exp}), and this
fact will be demonstrated below in Sec.~\ref{subsec:num_lin}. However,
it will also be demonstrated that (\ref{chi_1_exp}) becomes very
accurate in the spectral range of interest when $a\gtrsim 5{\rm nm}$.
Thus, the HRFR model yields a result for $\chi_1$, which is consistent
with the macroscopic limit.

The situation is dramatically different in the case of the nonlinear
susceptibility $\chi_3$. It follows from (\ref{eq:Dan}) that $\chi_3
\xrightarrow[]{a\rightarrow \infty} O(a^2)$. Therefore, there is no
``bulk'' limit for $\chi_3$. This is an unexpected result. While some
studies suggest that a positive correlation between $\chi_3$ and $a$
in a limited range of $a$ is consistent with experimental
measurements~\cite{drachev_04_2}, we can not expect this correlation
to hold for arbitrarily large values of $a$, as this would,
essentially, entail an infinite value of $\chi_3$ in bulk.  Such a
prediction appears to be unphysical. Of course, Rautian's theory is
not expected to apply to arbitrarily large values of $a$ because the
interaction potential (\ref{eq:V}) is written in the dipole
approximation and, moreover, it assumes that the electric field inside
the nanoparticle is potential, that is, $\nabla \times {\bf E}=0$ is a
good approximation. Still, the absence of a ``bulk'' limit for
$\chi_3$ is troublesome. We, therefore, wish to understand whether the
quadratic dependence of $\chi_3$ on $a$ is a property of the HRFR
model itself or an artifact of the additional approximations made in
deriving the analytical expression (\ref{eq:Dan}). More specifically,
we can state the following two hypotheses:
\begin{enumerate}
\item The quadratic dependence of $\chi_3$ on $a$ is an artifact of
  the approximations made in deriving the analytical expression
  (\ref{eq:Dan}) from (\ref{eq:chi3}) [these approximations are listed
  explicitly between Eqs.~(\ref{eq:B}) and (\ref{chi_1_an})]. In this
  case, we can expect that direct evaluation of (\ref{eq:chi3}) will
  not exhibit the quadratic growth.
\item The quadratic dependence of $\chi_3$ on $a$ is a property of the
  HRFR model itself. In particular, the absence of a ``bulk'' limit
  for $\chi_3$ can be caused by the following reasons: (i) The HRFR
  model neglects the retardation effects in large particles. (ii) The
  HRFR model does not account for the Hartree interaction potential.
  (In reality, interaction of the conduction electrons with the
  induced charge density may be important, especially, for computing
  nonlinear corrections.) (iii) The HRFR model makes use of a
  phenomenological boundary condition at the nanoparticle surface.
\end{enumerate}
Verification of these hypotheses was previously hindered by the
computational complexity of the problem. In what follows, we render
Rautian's theory amenable to direct numerical validation. Then we show
that the analytical approximation (\ref{eq:Dan}) is surprisingly good.
Therefore, the second hypothesis must be correct. 

\section{Rautian's theory further developed}
\label{sec:further}

It is possible to simplify~(\ref{eq:chi3}) without adopting any
approximations. To this end, we deviate from Rautian's approach of
using irreducible spherical tensors and $6j$-symbols. Instead, we
directly substitute the expressions
(\ref{eq:d_del}),(\ref{eq:del}),(\ref{b_R_def}) into (\ref{eq:chi3}).
We use the selection rules in (\ref{eq:del}) and the following
results:
\begin{subequations}
\begin{align}
& Z_l \equiv \sum_{m=-l}^l b^4_{lm}  = \frac{l(4l^2+1)}{15(4l^2-1)} \ , \\
& S_l \equiv \sum_{m=-l}^l b_{lm}^2 b_{l+1,m}^2 = \frac{2l(l+1)}{15(2l+1)}
\end{align}
\end{subequations}
to evaluate summations over all magnetic quantum numbers and over all
orbital quantum numbers but one. This leaves us with a five-fold
summation over four main quantum numbers and one orbital quantum
number. After some rearrangements, we arrive at the following
expression
\begin{equation}
\label{eq:Dnum}
\chi_3  = \frac{(ea)^4 A_{\rm at}^2}{(\hbar\omega)^3 \Omega} \sum_{l=1}^{\infty}
\left(Z_l P_l + S_l Q_l \right) \ ,
\end{equation}
\noindent
where
\begin{subequations}
\label{P_l_Q_l}
\begin{align}
P_l = \sum_{n_1,n_2,n_3,n_4} \left[ B_{n_3, l-1,n_4,l}^{n_1,l,n_2,l-1} +
  B_{n_1,l,n_3,l-1}^{n_2,l-1,n_4,l} \right] 
\nonumber \\
\times R_{n_3,l-1}^{n_1,l} R_{n_1,l}^{n_2,l-1} R_{n_2,l-1}^{n_4,l}
R_{n_4,l}^{n_3,l-1} \ , 
\label{P_l} \\
Q_l = \sum_{n_1,n_2,n_3,n_4} \left[ 
  B_{n_3,l-1,n_4,l}^{n_1,l,n_2,l+1} 
+ B_{n_1,l,n_3,l-1}^{n_2,l+1,n_4,l} \right. \nonumber \\
\left. 
+ B_{n_2,l+1,n_1,l}^{n_4,l,n_3,l-1}
+ B_{n_4,l,n_2,l+1}^{n_3,l-1,n_1,l} 
\right] \nonumber \\
\times 
R_{n_3,l-1}^{n_1,l} 
R_{n_1,l}^{n_2,l+1} 
R_{n_2,l+1}^{n_4,l} 
R_{n_4,l}^{n_3,l-1} \ .
\label{Q_l}
\end{align}
\end{subequations}
\noindent
This expression is exact within the HRFR model. The two-level
approximation corresponds to keeping only the first term in the
brackets in (\ref{eq:Dnum}) and, further, keeping only the terms with
$n_2=n_3$ and $n_1=n_4$ in (\ref{P_l}).

\section{Relating the internal and the applied fields}
\label{sec:Ei_Ee}

In the HRFR model, electrons move in a given, spatially-uniform
internal field (\ref{Ei_def}). In practice, one is interested in the
optical response of the nanoparticle to the external (applied) field.
We denote the amplitude of the external field by ${\bf A}_e = A_e
\hat{\bf z}$. The two fields differ because of a charge density
induced in the nanoparticle. The interaction of the conduction
electrons with the induced charge density is described by the Hartree
potential. However, rigorous introduction of the Hartree interaction
into the HRFR model is problematic. Doing so would require the
mathematical apparatus of density-functional theory. We can, however,
apply here the classical concept of the depolarizing field, although
this approach is less fundamental.

In the macroscopic theory, a sphere (either dielectric or conducting),
when placed in a spatially-uniform, quasistatic electric field of
frequency $\omega$ and amplitude ${\bf A}_e$, is polarized and
acquires a dipole moment of an amplitude ${\bf D}$. The electric field
inside the sphere is also spatially-uniform and has the amplitude
${\bf A}_i$. The induced charge accumulates at the sphere surface in a
layer whose width is neglected. Under these conditions, ${\bf A}_i =
{\bf A}_e - {\bf D}/a^3$. Note that a linear dependence between ${\bf
  D}$ and ${\bf A}_e$ is not assumed here. The form of the
depolarizing field, $-{\bf D}/a^3$, follows only from the assumption
of spatial uniformity of the internal field and from the usual
boundary conditions at the sphere surface.  Then the Hartree
interaction can be taken into account as follows.

Let us introduce the dimensionless variables $x=A_i/A_{\rm at}$ and
$y=A_e/A_{\rm at}$. Then we can expand $D$ in both variables:
\begin{subequations}
\label{D_Ai_Ae}
\begin{align}
\label{D_Ai}
& D = \Omega A_{\rm at} \left( \chi_1 x + \chi_3 x \vert x \vert^2 +
  \chi_5 x \vert x \vert^4 +
  \ldots \right) \ , \\
\label{D_Ae}
& D = \Omega A_{\rm at} \left( \alpha_1 y + \alpha_3 y \vert y \vert^2
  + \alpha_5 y \vert y \vert^4 +
  \ldots \right) \ ,
\end{align}
\end{subequations}
\noindent
where
\begin{equation}
\label{x_y}
x = y - \frac{D}{A_{\rm at}a^3} = y - \frac{4\pi}{3}\left( \alpha_1 y
  + \alpha_3 y \vert y \vert^2 + \ldots \right) \ .
\end{equation}
\noindent
Here we have accounted for the fact that there can be a phase shift
between the internal and the external fields; therefore, $A_i$ and
$A_e$ can not be real-valued {\em simultaneously}. In the theory
presented above, we assume that $A_i$ is real-valued, and this can
always be guaranteed by appropriately choosing the time origin. In
this case, $A_e$ is expected to be complex.

The coefficients $\chi_k$ in (\ref{D_Ai}) can be found from Rautian's
theory; our task is to find the coefficients $\alpha_k$ in
(\ref{D_Ae}) given the constraint (\ref{x_y}). To this end, we
substitute (\ref{x_y}) into (\ref{D_Ai}) and obtain a series in the
variable $y$. We then require that the coefficients in this series and
in (\ref{D_Ae}) coincide. This yields an infinite set of equations for
$\alpha_k$, the first two of which read
\begin{subequations}
\label{chi_alpha_eq}
\begin{align}
& \chi_1\left(1 - \frac{4\pi}{3}\alpha_1 \right) =
\alpha_1 \ , \\
& \chi_3 \left( 1 - \frac{4\pi}{3}\alpha_1 \right)
\left\vert 1 - \frac{4\pi}{3}\alpha_1 \right\vert^2  -
\chi_1 \frac{4\pi}{3} \alpha_3 = \alpha_3 \ .
\end{align}
\end{subequations}
\noindent
It is convenient to introduce the linear field enhancement factor
$f_1$ according to
\begin{equation}
\label{f1_def}
f_1 = \frac{1}{1 + (4\pi/3)\chi_1} = \frac{3}{\epsilon_1 + 2} \ ,
\end{equation}
\noindent
where $\epsilon_1 = 1 + 4\pi\chi_1$ is the linear dielectric
permittivity. Then the solutions to (\ref{chi_alpha_eq}) have the form
\begin{equation}
\label{chi_alpha_sol}
\alpha_1 = f_1 \chi_1 \ , \ \ \alpha_3 = f_1^2 \vert f_1\vert^2 \chi_3
\ .
\end{equation}
\noindent
The factor $f$, which relates the external and internal field
amplitudes according to $A_i = f A_e$, is then found from
\begin{equation}
\label{f_def}
f = \frac{x}{y} = 1 - \frac{4\pi}{3}\left(\alpha_1 + \alpha_3 \vert
  y\vert^2 + \ldots \right) \ .
\end{equation}
Using (\ref{chi_alpha_sol}), we find that, to first order in $I/I_{\rm
  at} \equiv \vert y \vert^2 = \vert A_e / A_{\rm at} \vert^2$,
\begin{equation}
\label{f_result}
f = f_1 - \frac{4\pi}{3}f_1^2 \vert f_1\vert^2 \chi_3 \frac{I}{I_{\rm at}} \
.
\end{equation}
\noindent
Here we have introduced the intensity of the incident beam, $I =
(c/2\pi)\vert A_e\vert^2$, and the ``atomic'' intensity $I_{\rm at} =
(c/2\pi)\vert A_{\rm at} \vert^2$.

Note that our approach to finding the field enhancement factor $f$ is
somewhat different from that adopted in
Ref.~\onlinecite{drachev_04_2}, where the expansions (\ref{D_Ai}) has
been truncated at the third order and the truncated expression was
assumed to be exact. The results obtained in the two approaches
coincide to first order in $I/I_{\rm at}$. In
Ref.~\cite{drachev_04_2}, higher order corrections to this result have
also been obtained. In our approach, these corrections depend on the
higher-order coefficients $\chi_5$, $\chi_7$, etc., which have not
been computed by Rautian.

We finally note that the phenomenological accounting for the Hartree
interaction described in this section, while is necessary for
comparison with the experiment, does not remove the two main
difficulties of the HRFR model. Specifically, it does not fix the
low-frequency limit for $\chi_1$ and does not affect the $\propto a^2$
dependence of $\chi_3$. Regarding the low-frequency limit, we note
that $\lim_{\omega\rightarrow 0}f = 0$ and the internal field in the
nanoparticle tends to zero in this limit. The induced macroscopic
charge is localized at the sphere surface where the electric field
jumps abruptly. In a more accurate microscopic picture, the width of
this surface layer is finite and the electric field changes smoothly
over the width of this layer. Unfortunately, the classical concept of
depolarizing field can not capture surface phenomena of this kind.

\section{Numerical results}
\label{sec:num}

\subsection{Convergence}

We have computed the Bessel function zeros $\xi_{nl}$ using the method
of bisection and achieved a numerical discrepancy of the equation
$j_l(\xi_{nl})=0$ of less than $10^{-15}$ for all values of indices.
Since the function $j_l(x)$ is approximately linear near its roots, we
believe that we have computed $\xi_{nl}$ with sufficiently high
precision.

\begin{figure}
\noindent\hspace*{-1cm}\input{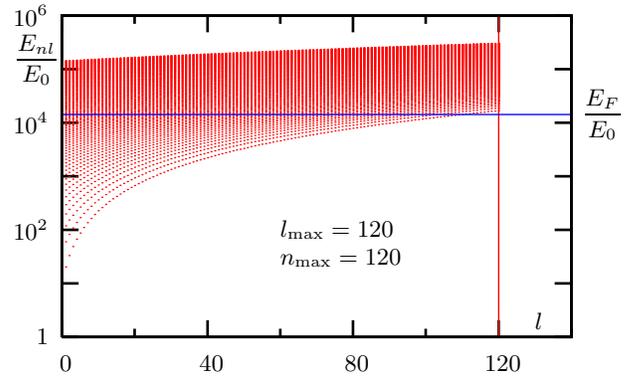}
\caption{\label{fig:1} (color online) Energy eigenstates, which enter the
  summation according to (\ref{eq:Dnum}),(\ref{P_l_Q_l}), for
  $a=10{\rm nm}$ and $l_{\rm max}=n_{\rm max}=120$. The horizontal
  blue line shows the Fermi energy. In this example, the total number
  of states below the Fermi surface is ${\mathscr N} \approx 2.4\cdot
  10^5$ (counting all degeneracies) and the total number of states
  shown in the figure is $2n_{\rm max}l_{\rm max}(l_{\rm max}+2)
  \approx 3.5\cdot 10^6$.}
\end{figure}

The summation over $l$ in (\ref{eq:Dnum}) was truncated so that $l\leq
l_{\rm max}$ and the quadruple summation in (\ref{P_l_Q_l}) was
truncated so that $n_1,n_2,n_3,n_4\leq n_{\rm max}$. A typical set of
energy levels used in the summation is shown in Fig.~\ref{fig:1} for
the case $a=10{\rm nm}$, $l_{\rm max}=n_{\rm max}=120$. Here the
energy levels (normalized to $E_0$) are shown by dots and the
horizontal axis corresponds to the orbital number, $l$.  Referring to
Fig.~\ref{fig:1}, we note that $l_{\rm max}$ has been chosen so that
all states with $l \geq l_{\rm max}$ are above the Fermi surface.
Since the electron transitions occur between two states with $l$ and
$l^\prime$ such that $l^\prime = l \pm 1$, the factors $N_{\mu\nu}$
for any transition involving the states with $l \geq l_{\rm max}$ are
zero (or exponentially small at finite temperatures).  It can be seen
that convergence with $l$ is very fast -- contribution of the terms in
(\ref{eq:Dnum}) with $l \geq l_{\rm max}$ is either zero (at $T=0$) or
exponentially small.

The choice of $n_{\rm max}$ is a more subtle matter. Since there are
no selection rules on $n$, transitions can occur between two states
(one below and one above the Fermi surface) with very different values
of $n$ and, correspondingly, very different energies. However,
transitions with energy gaps much larger than $\hbar\omega$ are
suppressed by the Lorentzian factors (\ref{Lambda_def}). In most
numerical examples, we have chosen $n_{\rm max}$ so as to account for,
at least, all transitions with the energy gaps of $\Delta E \leq
3\hbar\omega$. Many (but not all) transitions with larger energy gaps
were also accounted for. This approach yields a result with seven
significant figures. However, it results in too many terms in the
summation when $\hbar \omega \sim E_F$ and $a\geq 10{\rm nm}$. For
these values of parameters, we have used a smaller $n_{\rm max}$ so as
to account for, at least, all transitions with $\Delta E \leq \hbar
\omega$. We estimate that the relative error incurred by this
truncation is $\lesssim 10\%$.

\subsection{Linear response}
\label{subsec:num_lin}

We begin by considering the linear susceptibility $\chi_1$. In
computations, we use the commonly accepted parameters for silver,
$\hbar\omega_p = 8.98{\rm eV}$ ($\lambda_p = 2\pi c/\omega_p = 138{\rm
  nm}$) and $\gamma_\infty/\omega_p = 0.002$. Here the relaxation
constant $\Gamma_2$, which enters (\ref{eq:chi1}), is determined from
$\gamma_\infty = 2\Gamma_2$ (see (\ref{gamma_a})). The frequencies
used satisfy the condition $\gamma_\infty/\omega \ll 1$. More
specifically, the ratio $\omega/\omega_p$ varies in the range $0.02
\leq \omega/\omega_p \leq 1$. We do not consider the frequencies above
$\omega_p$ because silver exhibits strong interband absorption in that
spectral range. Except when noted otherwise, all computations have
been carried out at $T=0$.

\begin{figure}
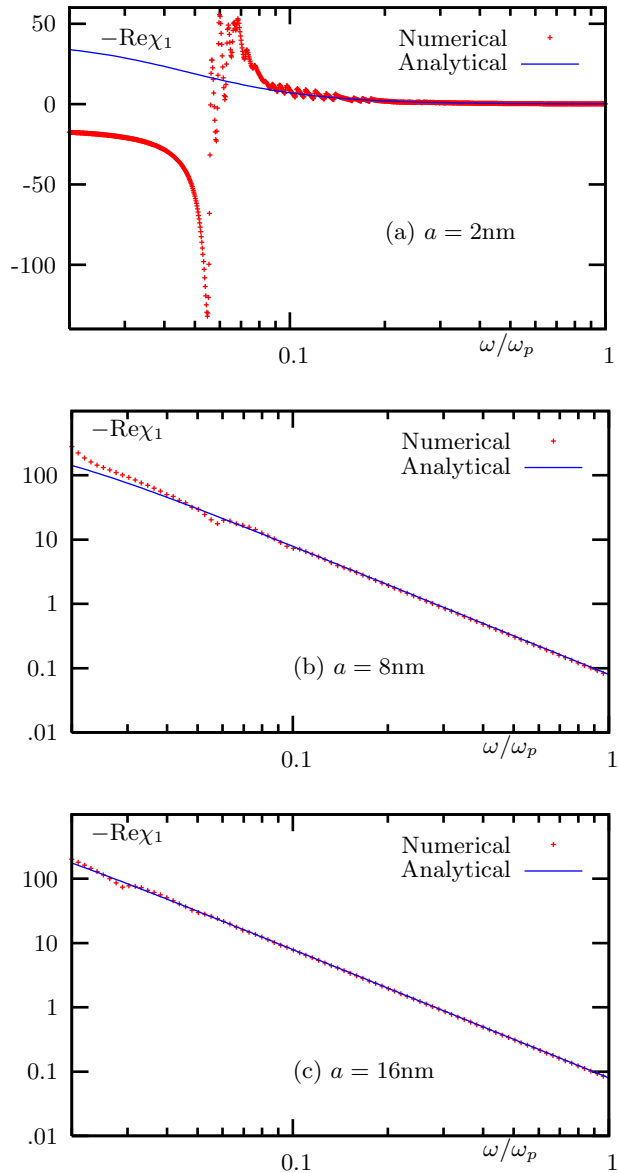

\noindent\hspace*{-1.0cm}\input{fig2a.tex}
\noindent\hspace*{-1.1cm}\input{fig2b.tex}
\noindent\hspace*{-1.1cm}\input{fig2c.tex}
\caption{\label{fig:2} (color online) The quantity $-{\rm Re}\chi_1$ as a
  function of frequency for particles of different radius $a$, as
  labeled. Centered symbols correspond to direct numerical evaluation
  of (\ref{eq:chi1}) and continuous curves show the Drude formula
  (\ref{chi_1_D}) in which the size-corrected relaxation constant
  $\gamma$ (\ref{gamma_a}) has been used.}
\end{figure}

Fig.~\ref{fig:2} displays the quantity $-{\rm Re}\chi_1$ computed
numerically by direct evaluation of (\ref{eq:chi1}) and by the Drude
formula (\ref{chi_1_D}) with the size-corrected relaxation constant
$\gamma$ (\ref{gamma_a}). The factor $g_1$ in (\ref{gamma_a}) has been
computed using the analytical formula (\ref{g1_eval}). At sufficiently
high frequencies, the Drude model predicts that $-4\pi {\rm Re}\chi_1
\approx (\omega_p/\omega)^2$ and this behavior is reproduced for all
radiuses considered with good precision. However, at smaller
frequencies, there are differences between the analytical
approximation and the numerical results. These differences are
especially apparent for $a=2{\rm nm}$. In this case, the optical
response of the sphere is, effectively, dielectric rather than
metallic for $\omega \lesssim 0.06\omega_p$. A similar behavior has
been observed at $a=4{\rm nm}$ (data not shown). The emergence of a
dielectric response in metal nanoparticles of sufficiently small size
at sufficiently low frequencies has been overlooked in the past. It
occurs due to discreteness of the energy states.  Consider a particle
with $a=2{\rm nm}$ at zero temperature. In this case, the
lowest-energy electronic transition, which is allowed by Fermi
statistics (that is, a transition with $N_{\mu\nu}\neq 0$), occurs
between the states $(n=1,l=18)$ and $(n^\prime=1,l^\prime=19)$.  The
corresponding transition frequency is $\omega_{\rm min} \approx
0.056\omega_p$. It can be seen from Fig.~\ref{fig:2}(a) that the
particle becomes dielectric for $\omega \lesssim \omega_{\rm min}$.

\begin{figure}
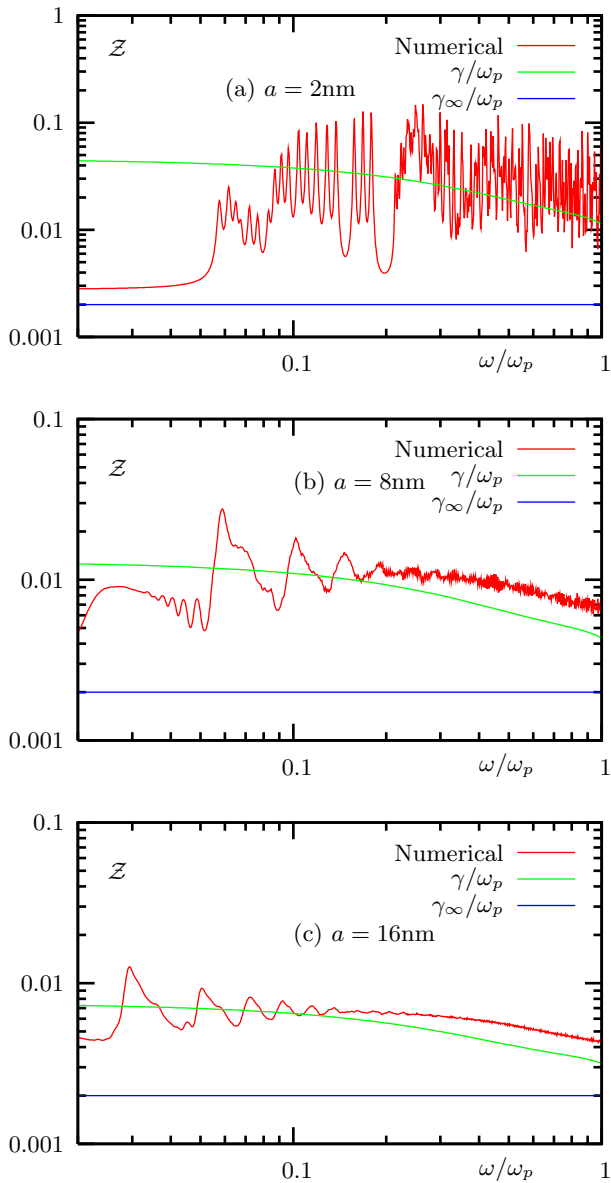

\noindent\hspace*{-1.0cm}\input{fig3a.tex}
\noindent\hspace*{-1.0cm}\input{fig3b.tex}
\noindent\hspace*{-1.0cm}\input{fig3c.tex}
\caption{\label{fig:3} (color online) The quantity ${\mathcal Z}$ defined in
  (\ref{Zet_def}) as a function of frequency for different particle
  radiuses, as labeled. Results of direct numerical computation are
  compared to the size-corrected Drude relaxation constant $\gamma$
  (given by (\ref{gamma_a})) and to its bulk value $\gamma_\infty$
  (obtained in the limit $a\rightarrow \infty$).}
\end{figure}

We now turn to consideration of the relaxation phenomena. To this end,
we plot in Fig.~\ref{fig:3} the quantity
\begin{equation}
\label{Zet_def}
{\mathcal Z} = -\frac{1}{4\pi}\frac{\omega_p}{\omega} {\rm Im}
\frac{1}{\chi_1}
\end{equation}
\noindent
as a function of frequency. We note that ${\mathcal Z}$, as defined in
(\ref{Zet_def}), is positive for all passive materials and, in the
Drude model, ${\mathcal Z} = \gamma/\omega_p$; here $\gamma$ is
size-corrected. It can be seen that the analytical formula
(\ref{gamma_a}) captures the relaxation phenomena in the nanoparticle
surprisingly well. However, as in Fig.~\ref{fig:2}(a), the analytical
approximation breaks down when $a=2{\rm nm}$ and $\omega \lesssim
0.06\omega_p$. A similar breakdown was observed for $a=4{\rm nm}$
(data not shown). For all other values of parameters, the numerically
computed ${\mathcal Z}$ is reasonably close to the size-corrected
value of $\gamma/\omega_p$ and exhibits the same overall behavior. The
small systematic error at higher frequencies is, most likely, caused
by the approximation (\ref{g1_def}) for $g_1$. It was, in fact,
mentioned by Rautian that (\ref{g1_def}) is hardly accurate when
$\hbar \omega \sim E_F$.

The fine structure visible in Fig.~\ref{fig:3}(a,b) is due to
discreteness of electron states. The allowed transition frequencies
can be ``grouped'', which results in the appearance of somewhat
broader peaks, clearly seen in Fig.~\ref{fig:3}(b,c). While spectral
signatures of discrete states in metal nanoparticles have been
observed experimentally (including the effect of
``grouping'')~\cite{drachev_04_3}, the positions of individual
spectral peaks should not be invested with too much significance. In
any realistic system, these peaks will be smoothed out by particle
polydispersity, variations in shape, and by nonradiative relaxation
and energy transfer to the surrounding medium.

The finite-size correction (\ref{gamma_a}) to the Drude relaxation
constant is widely known and used. However, the derivations of
(\ref{gamma_a}) have been, so far, either heuristic or relied on
poorly controlled approximations. In Fig.~\ref{fig:3}, we have
provided, to the best of our knowledge, the first direct,
first-principle numerical verification of (\ref{gamma_a}) and of its
limits of applicability.

\subsection{Nonlinear response}
\label{subsec:num_nonl}

We next turn to the nonlinear susceptibility $\chi_3$. The same
parameters for silver as before will be used. In addition, the
calculations require the relaxation constant $\Gamma_1$. As was
mentioned above, the experimental value of $\Gamma_1$ can not be
inferred by observing the linear optical response. It was previously
suggested~\cite{drachev_04_2} that $\Gamma_2/\Gamma_1 \approx 10$.
This value will be used below.

\begin{figure}
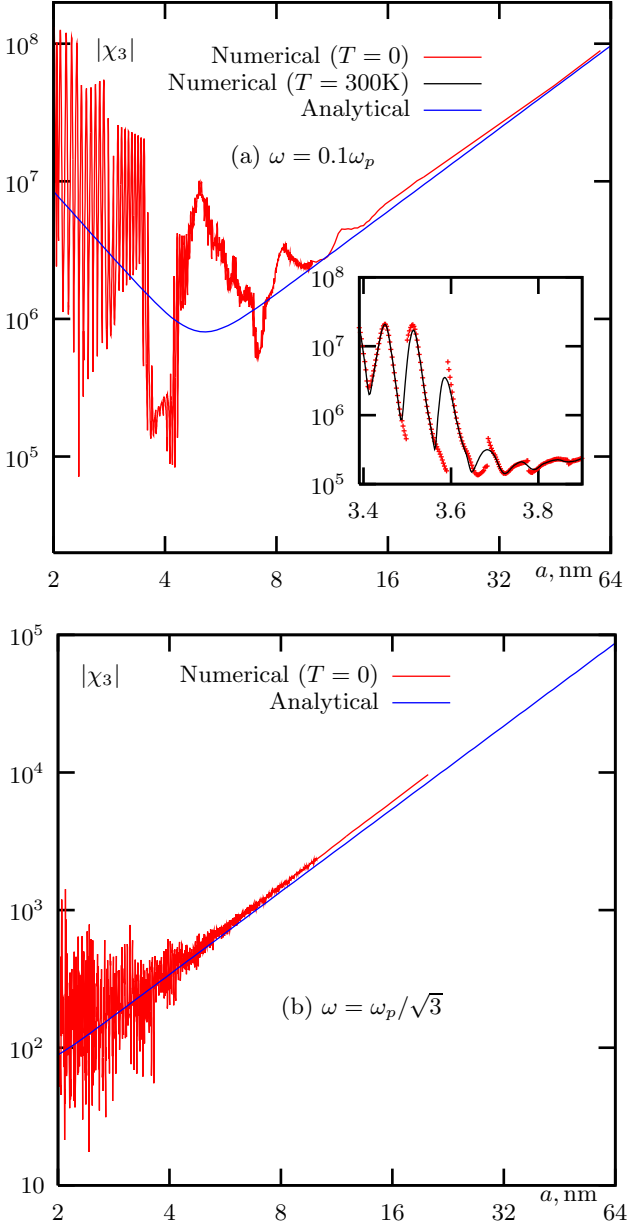

  \hspace*{-1cm}\noindent\input{fig4_a.tex}
  \hspace*{-1cm}\noindent\input{fig4_b.tex}
\caption{\label{fig:4} (color online) Absolute value of the nonlinear
  susceptibility, $\vert \chi_3 \vert$, computed by direct evaluation
  of (\ref{eq:Dnum}) and by analytical approximation (\ref{eq:Dan}) as
  a function of the particle radius for $\omega=0.1\omega_p$ (a) and
  for $\omega = \omega_p/\sqrt{3} \approx 0.58\omega_p$ (b).
  Logarithmic scale is used on both axes. The inset in panel (a) shows
  a zoom of the plot for $3.4{\rm nm} \leq a \leq 3.7{\rm nm}$. In the
  inset, the results of evaluating (\ref{eq:Dnum}) at $T=0$ and at
  $T=300K$ are shown.}
\end{figure}

In Fig.~\ref{fig:4}, we plot the absolute value of $\chi_3$ as a
function of the particle radius $a$ for $\omega=0.1\omega_p$ and for
the Frohlich frequency $\omega = \omega_p/\sqrt{3} \approx
0.58\omega_p$, and compare the results of direct numerical evaluation
of (\ref{eq:Dnum}) to the analytical approximation (\ref{eq:Dan}). In
the case $\omega=0.1\omega_p$, the analytical approximation is very
accurate for $a \gtrsim 8{\rm nm}$ and gives the correct overall trend
for $a\lesssim 4{\rm nm}$. A systematic discrepancy of unknown origin
between the approximate and the numerical results is observed for
$4{\rm nm} < a < 8{\rm nm}$. In the case $\omega=\omega_p/\sqrt{3}$,
the analytical approximation gives the correct trend in the whole
range of $a$ considered. Note that, in the case $\omega=0.1\omega_p$,
the absolute value of $\chi_3$ is dominated by ${\rm Re}\chi_3$ and
${\rm Im}\chi_3$ for large and small values of $a$, respectively. When
$\omega=\omega_p/\sqrt{3}$, the real part of $\chi_3$ is dominating
for all values of $a$ used in the figure.

Consider first particles with $a\lesssim 4{\rm nm}$. As expected, the
discreteness of energy levels plays an important role in this case and
results in a series of sharp maxima and minima of $\vert \chi_3(a)
\vert$. As is shown in the inset of Fig.~\ref{fig:4}(a), the function
$\vert \chi_3(a) \vert$ is discontinuous. These discontinuities are
artifacts of the zero temperature approximation. The introduction of a
finite temperature ($T=300K$) removes the discontinuities (see the
inset) but does not eliminate the fine structure of the curve. Note,
however, that the computations have been carried out with a very fine
step in $a$, which is, arguably, unphysical: the parameter $a$ in a
real nanosphere can change only in quantized steps of the order of the
lattice constant $h$ ($\approx 0.41{\rm nm}$ in silver).  Moreover,
the fine structure of $\chi_3$ is unlikely to be observable
experimentally due to the unavoidable effects of particle
polydispersity.  Therefore, the general trend given by the analytical
approximation (\ref{eq:Dan}) can be a more realistic estimate of
$\chi_3$ for $a\lesssim 4{\rm nm}$.

Next consider the large-$a$ behavior. For $a\gtrsim 8{\rm nm}$, the
analytical and the ``exact'' formulas yield results, which are
scarcely distinguishable. In particular, the quadratic growth of
$\chi_3$ with $a$ has been confirmed up to $a=64{\rm nm}$ in the case
$\omega = 0.1\omega_p$ -- the largest radius for which numerical
evaluation of (\ref{eq:Dnum}) is still feasible. This confirms
Hypothesis 2 stated above, namely, that the quadratic growth of
$\chi_3(a)$ is a property of the HRFR model itself rather than of the
additional approximations, which were made to derive the analytical
results.

\begin{figure}
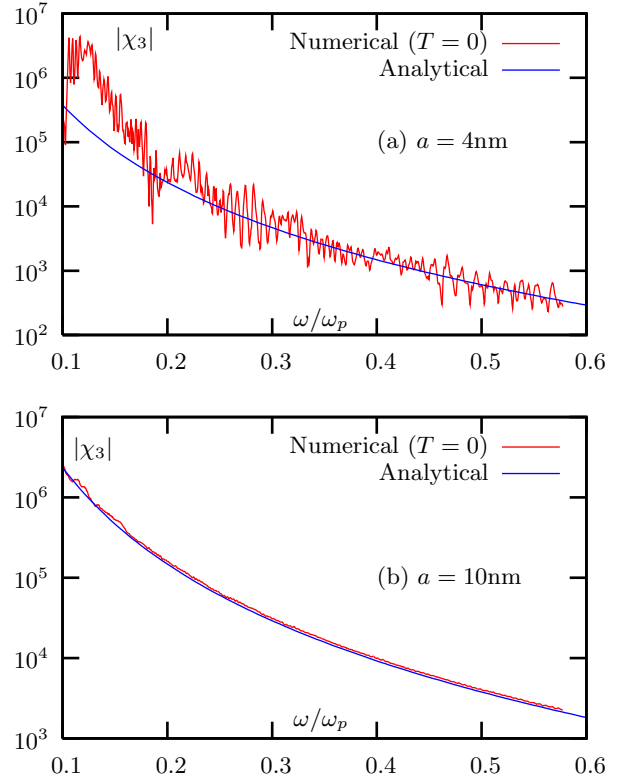

  \hspace*{-1cm}\noindent\input{fig5_a.tex}
  \hspace*{-1cm}\noindent\input{fig5_b.tex}
\caption{\label{fig:5} (color online) Absolute value of the nonlinear
  susceptibility, $\vert \chi_3 \vert$, computed by direct evaluation
  of (\ref{eq:Dnum}) at zero temperature and by analytical
  approximation (\ref{eq:Dan}) as a function of the frequency for
  $a=4{\rm nm}$ (a) and for $a=10{\rm nm}$ (b).}
\end{figure}

In Fig.~\ref{fig:5}, we study the dependence of $\vert \chi_3 \vert$
on the frequency $\omega$ for fixed values of $a$. It can be seen that
the accuracy of Rautian's approximation improves for larger particles
and higher frequencies. At $a=10{\rm nm}$, the approximation is nearly
perfect in the full spectral range considered.

\subsection{Magnitude of the nonlinear effect and comparison with the
  classical theory of electron confinement}
  
In the previous subsection, we have plotted the coefficient $\chi_3$,
which appears in the expansion (\ref{eq:D}). The dimensionless
parameter of this expansion, $A_i/A_{\rm at}$, contains the amplitude
of the internal electric field, $A_i$. However, it is the amplitude of
the external (applied) field, $A_e$, which can be directly controlled
in an experiment. The incident beam intensity is given by $I =
(c/2\pi) \vert A_e \vert^2$. We can use the results of
Sec.~\ref{sec:Ei_Ee} to write
\begin{equation}
\label{D_Q}
D = \Omega A_e \left[\alpha_1 + \alpha_3 (I/I_{\rm at}) + \ldots \right] \ ,
\end{equation}
where $\alpha_1$ and $\alpha_3$ are related to $\chi_1$ and $\chi_3$
by (\ref{chi_alpha_sol}) and $I_{\rm at} = (c/2\pi) A_{\rm at}^2$ is
the characteristic ``atomic'' intensity. The quantity $I_{\rm at}$ can
be expressed in terms of the fundamental physical constants and the
material-specific parameter $\ell$. In the case of silver, $I_{\rm at}
\approx 2.3\cdot 10^{14}{\rm W/cm}^2$. Obviously, intensities of such
magnitude are not achievable in any experiment. However, the magnitude
of the nonlinear correction can be amplified by the two important
effects~\cite{boyd_book_92}: the effect of synchronism (constructive
interference), which is not considered here, and by the effect of
local field enhancement, which will be taken into account by using the
expressions derived in Sec.~\ref{sec:Ei_Ee}.

We will also compare the expression (\ref{D_Q}) to the results
obtained from the purely classical arguments~\cite{panasyuk_08_1}. In
Ref.~\onlinecite{panasyuk_08_1}, we have argued the surface charge in
a polarized metal nanoparticle can not be confined to an infinitely
thin layer. When the width of this layer is not negligible (compared
to the particle radius), a nonlinear correction to the particle
polarizability is obtained. After some rearrangement of the formulas
derived in this reference, we can express the amplitude $D$, defined
analogously to (\ref{eq:dwt}), as
\begin{equation}
\label{D_C}
D = \Omega A_e \left[\beta_1 + \beta_3 \sqrt{I/I_{\rm at}} + \ldots\right] \ ,
\end{equation}
where 
\begin{subequations}
\label{beta}
\begin{align}
\label{beta_1}
\beta_1 & = \frac{3}{4\pi}\frac{\omega_p^2/3}{\omega_p^2/3 - \omega^2 -
  i\gamma\omega} \\
\label{beta_3}
\beta_3 & =  - \frac{3}{\pi}\frac{\ell}{a} \beta_1 \vert \beta_1 \vert \ ,
\end{align}
\end{subequations}
\noindent
We note that $\alpha_1 \xrightarrow[]{\omega/\gamma \rightarrow
  \infty} \beta_1$. That is, the linear polarizabilities of both
theories are the same in the region of parameters where the theories
are applicable. The classical theory, however, does not contain the
low frequency anomaly in the linear polarizability. On the other hand,
relaxation is introduced in Ref.~\onlinecite{panasyuk_08_1} through the
phenomenological parameter $\gamma$ whose dependence on $a$ can not be
deduced theoretically. Below, we use the result of the quantum theory,
namely, Eq.~(\ref{gamma_a}) for the relaxation constant $\gamma$
(\ref{beta_1}).

It can be seen that the classical and quantum expressions for $D$ are
quite different. The first non-vanishing nonlinear correction in
(\ref{beta_3}) is of the order of $\sqrt{I/I_{\rm at}}$ but contains
an additional small parameter $\ell/a$. Thus, the nonlinear correction
depends differently on the incident intensity, frequency and the
particle radius $a$ in the two theories. Additionally, Rautian's
theory contains the parameter $\Gamma_1/\Gamma_2$, which does not
enter into the classical theory. These factors complicate a direct
comparison of the two results. We will, therefore, focus on the trends
for one particular value of the incident power, $I=10{\rm
  kW/cm^2}$. One should bear in mind that the nonlinear corrections
depend on the incident power differently in the two theories.

\begin{figure}
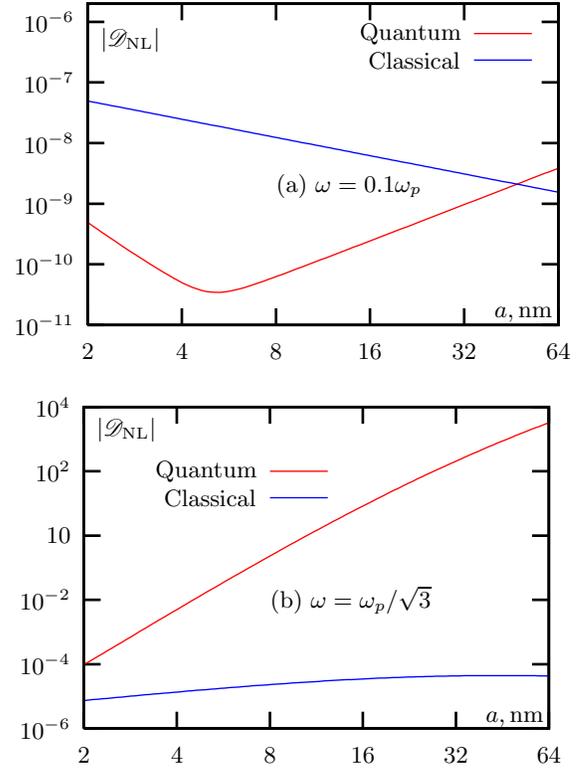

  \hspace*{-1cm}\noindent\input{fig6_a.tex}
  \hspace*{-1cm}\noindent\input{fig6_b.tex}
\caption{\label{fig:6} (color online) Absolute value of the nonlinear correction to the
  nanoparticle polarizability, ${\mathscr D}_{\rm NL}$, computed using
  Eq.~(\ref{D_Q}) (the ``quantum'' curves) and Eq.~(\ref{D_C}) (the
  ``classical'' curves) for $\omega=0.1\omega_p$ (a) and
  $\omega=\omega_p/\sqrt{3} \approx 0.58\omega_p$ (b). Here ${\mathscr
    D}_{\rm NL} = \alpha_3 (I/I_{\rm at})$ (the ``quantum'' curves)
  and ${\mathscr D}_{\rm NL} = \beta_3 \sqrt{I/I_{\rm at}}$ (the
  ``classical'' curves). The incident power is $I=10{\rm kW/cm^2}$,
  $I/I_{\rm at} \approx 4.3 \cdot 10^{-11}$. To compute $\alpha_3$,
  the internal field enhancement factor has been taken into account
  according to (\ref{chi_alpha_sol}).}
\end{figure}

In Fig.~\ref{fig:6}, we plot the absolute value of the nonlinear
correction to the particle polarizability normalized by its volume as
a function of radius for the same values of frequency as were used in
Fig.~\ref{fig:4}. We denote the quantity being plotted by ${\mathscr
  D}_{\rm NL}$ and 
\begin{equation}
\label{math-D_def}
{\mathscr D}_{\rm NL} \equiv \left\{
\begin{array}{ll}
\alpha_3 (I/I_{\rm at})    \ , & \mbox{in the ``quantum''
case} \\
\beta_3\sqrt{I/I_{\rm at}} \ , & \mbox{in the ``classical'' case} \ .
\end{array}
\right.
\end{equation}
\noindent
The nonlinear effects should be observable in measurements with
incoherent light if $\vert {\mathscr D}_{\rm NL} \vert \gtrsim 1$. If
$\vert {\mathscr D}_{\rm NL} \vert \ll 1$, detection of the nonlinear effects
requires coherent laser excitation and utilization of the effect of
synchronism.

The parameters used in Fig.~\ref{fig:6} are such that the approximate
analytical formulas for $\chi_1$ (\ref{chi_1_D}) and $\chi_3$
(\ref{eq:Dan}) are fairly accurate, as was demonstrated above.
Correspondingly, we have used these formulas to generate the curves,
which are displayed in Fig.~\ref{fig:6}. To obtain the ``quantum''
curves, the following procedure has been followed. First, we have
computed the function $\chi_3(a)$ according to (\ref{eq:Dan}) for each
frequency considered.  Then we have computed $\chi_1(a)$ according to
(\ref{chi_1_D}) for the same frequencies. In Eq.~(\ref{chi_1_D}), we
have accounted for the dependence of the relaxation constant $\gamma$
on $a$ according to (\ref{gamma_a}). The computed function $\chi_1(a)$
was used to compute the linear field enhancement factor $f_1(a)$
according to (\ref{f1_def}). Finally, we have used the functions
$f_1(a)$ and $\chi_3(a)$ to compute $\alpha_3(a)$ according to
(\ref{chi_alpha_sol}). The result was multiplied by $I/I_{\rm at}
\approx 4.3 \cdot 10^{-11}$. In the ``classical'' case, $\beta_3$ was
computed according to (\ref{beta}), where the relaxation constant
$\gamma$ was size-corrected according to (\ref{gamma_a}).

We now discuss the curves shown in Fig.~\ref{fig:6} in more detail.
First, in the ``quantum'' case, ${\mathscr D}_{\rm NL}$ exhibits an
unlimited growth with $a$ when $a\rightarrow \infty$. In the classical
case, this growth is suppressed.  As can be seen, the ``classical''
${\mathscr D}_{\rm NL}$ decreases with $a$ in the case
$\omega=0.1\omega_p$ and seems to reach a finite limit in the case
$\omega=\omega_p/\sqrt{3}$.  In reality, however, the ``classical''
curve in Fig.~\ref{fig:6}(b), reaches a maximum at $a\approx 44{\rm
  nm}$ and then slowly approaches zero (the range of radiuses, which
is necessary to see this behavior clearly, is not shown in the
figure). In the classical theory, the nonlinearity is an effect of the
finite size, which vanishes in the limit $a\rightarrow \infty$.

Second, when $\omega=0.1\omega_p$, the local-field enhancement factor
in the ``quantum'' theory is $\vert f_1 \vert^4 \sim 10^{-8}$. That
is, the field is effectively screened in the interior of the
nanoparticle.  Correspondingly, the nonlinear effect is very weak. In
the ``classical'' theory, the field enhancement factor is different,
namely, it is $\vert \beta_1 \vert^2 \sim 1$.  This dramatic
difference is explained by the fact that the classical theory
considers the induced electron density near the nanoparticle surface
where the electric field is not entirely screened. At the Frohlich
frequency, $\omega=\omega_p/\sqrt{3}$, the situation is quite
different: we have
\begin{equation*}
\vert f_1\vert ^4 \sim \omega_p^4/9\gamma^4
\xrightarrow[]{a\rightarrow \infty} \omega_p^4/9\gamma_\infty^4\approx
7\cdot 10^9 \ .
\end{equation*}
\noindent
Correspondingly, the ``quantum'' nonlinear correction can become very
large at the Frohlich frequency; this is illustrated in
Fig.~\ref{fig:6}(b). This result is probably unphysical -- one can not
expect that ${\mathscr D}_{\rm NL} \sim 10^3$ at the modest incident
intensity of $10{\rm kW/cm^2}$. The classical curve, however, is still
bounded at the Frohlich frequency below $10^{-4}$. We can conclude
therefore that the local-field correction plays a disproportionate
role in the quantum theory and that, if used unscrupulously, it can
predict utterly unrealistic magnitudes of the nonlinear effect.

\section{Summary of findings and discussion}
\label{sec:summ}

In this paper, we have further developed the quantum theory of
Refs.~\onlinecite{hache_86_1,rautian_97_1} (the HRFR model). The goal
was to describe the frequency and size dependence of linear and
nonlinear optical susceptibilities of spherical metal nanoparticles.
We have used the HRFR model without modification but have managed to
simplify the previously published expressions to a point where these
expressions became amenable to direct numerical implementation. Then,
we have computed the linear and nonlinear susceptibilities numerically
for various frequencies and various particle sizes and compared the
obtained results to Rautian's analytical approximations.  Previously,
numerical computations of this kind have been hindered by the
overwhelming computational complexity of the problem. We have also
compared the predictions of the quantum theory of size-dependent
optical susceptibilities with the predictions of a purely classical
theory of Ref.~\onlinecite{panasyuk_08_1}. The following findings can
be reported:

\begin{enumerate}
\item We have found that the approximate formulas derived by
  Rautian~\cite{rautian_97_1} are surprisingly accurate in a wide
  range of parameters despite the use of a number of approximations.
  In particular, we have, for the first time, verified from first
  principles the correctness of the widely-used finite-size correction
  to the Drude relaxation constant (\ref{gamma_a}).
\item We have found that, for sufficiently small values of radius and
  frequency, Rautian's approximations break down due to the
  discreteness of electron energy levels. At sufficiently small
  frequencies, a silver particle with $a \lesssim 4{\rm nm}$ in radius
  behaves as a dielectric. This effect is illustrated in
  Fig.~\ref{fig:2}(a) for $a=2{\rm nm}$.
\item We have found that phenomenologically accounting for the
  local-field correction (see Sec.~\ref{sec:Ei_Ee} for details) does
  not remove the two main difficulties, which are encountered in the
  HRFR model, namely, the incorrect small-$\omega$ asymptote for the
  linear susceptibility $\chi_1$ and the absence of a ``bulk'' limit
  for the nonlinear susceptibility $\chi_3$. It appears that obtaining
  the correct asymptotes requires the rigorous account for the Hartree
  interaction potential. It is also conceivable that obtaining the
  correct large-$a$ asymptote requires accounting for the retardation
  effects.  However, the classical theory of
  Ref.~\onlinecite{panasyuk_08_1} is quasistatic but does not possess
  a large-$a$ anomaly. This suggests that the main focus in further
  development of Rautian's theory should be on a more accurate
  inclusion of Hartree interaction.
\end{enumerate}

One additional comment on the theory developed here are necessary.
First, we have computed only a particular case of the nonlinear
susceptibility $\chi^{(3)}(\omega;\omega_1,\omega_2,\omega_3)$. More
specifically, the coefficient $\chi_3$ defined in (\ref{eq:D}) is
related to the latter quantity by $\chi_3 = A_{\rm at}^{-2}
\chi^{(3)}(\omega;\omega,-\omega,\omega)$. However, consideration of
transient processes, generation of combination frequencies and
harmonics requires the knowledge of
$\chi^{(3)}(\omega;\omega_1,\omega_2,\omega_3)$ as a function of all
of its arguments. This is an important consideration. High incident
intensities are usually obtained in short laser pulses. Moreover, many
modern photonics applications such as waveguiding, etc., utilize short
wave-packets. Therefore, a proper description of optical
nonlinearities in a transient process is very important. Generalizing
the mathematical formalism described in this work to include three
independent frequencies is not conceptually difficult, although can
lead to cumbersome calculations.

In summary, the HRFR model forms a perfect theoretical framework for
studying optical nonlinearities and finite-size effects in
nanoparticles. The only viable alternative to using this model is to
resort to density-functional theory (DFT). In a recent
paper~\cite{panasyuk_11_1}, we have applied DFT to study the nonlinear
electromagnetic response of metal nanofilms, but only at very low
frequencies, well below plasmonic resonance of the system, and
neglecting the relaxation phenomena. Higher frequencies, which are of
interest in plasmonics, can be studied with the use of time-dependent
DFT (TDDFT). Although TDDFT has been used successfully to compute
linear response of
nanoparticles~\cite{ekardt_84_1,lerme_99_1,vasiliev_02_1}, and, in
particular, to study the effects of surface adsorption of various
molecules on the relaxation phenomena in
metal~\cite{zhu_08_1,gavrilenko_10_1}, the difficulties here are
formidable. Most importantly, there is almost no hope of obtaining
analytical approximations within DFT. It appears, therefore, that
devising a way to include the Hartree interaction potential in the
master equation (\ref{eq:rhoDifEq}) would be a useful and
practically-relevant development of the HRFR model and of Rautian's
theory. Perhaps, some elements of DFT can be used to achieve this.

This work was supported by the NSF under the Grant No. DMR0425780.  One of
the authors (GYP) is supported by the National Research Council Senior
Associateship Award at the Air Force Research Laboratory.

\bibliographystyle{apsrev}
\bibliography{abbrev,book,master,rautian,local}

\appendix

\section{Functions $g_1(\kappa)$ and $g_3(\kappa)$}

\begin{figure}
\noindent\hspace*{-1cm}\input{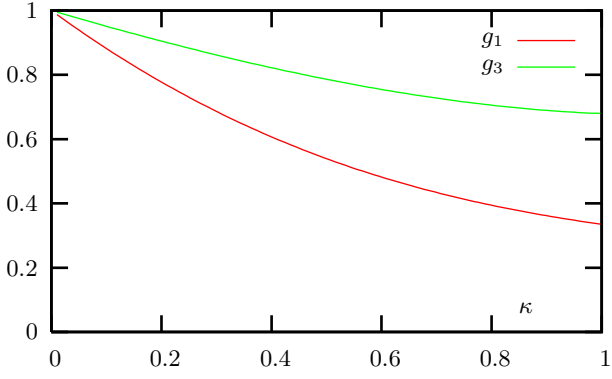}
\caption{\label{fig:g1_g3} (color online) Functions $g_1(\kappa)$ and
  $g_3(\kappa)$.} 
\end{figure}

The integrals (\ref{g1_def}) and (\ref{g3_def}) can be evaluated
analytically with the following results:
\begin{align}
\label{g1_eval}
\kappa g_1(\kappa) = -\frac{2}{9} + \frac{6\kappa}{7} -
\frac{6\kappa^2}{5} + \frac{2\kappa^3}{3}
+ \frac{2\sqrt{1+\kappa}}{315} \nonumber \\
\times \left(35 + 5\kappa - 6\kappa^2 + 8\kappa^3 - 16\kappa^4\right)
\ ,
\end{align}
\begin{align}
\label{g3_eval}
640 \kappa g_3(\kappa) 
= & \left[ 128 + \kappa (\kappa + 2) (88 + 5 \kappa (3 \kappa - 8) ) \right] \sqrt{1 + \kappa} \nonumber 
\\
- & \left[ 128 + \kappa(\kappa - 2) 
(168 - 5 \kappa (3 \kappa + 8)) \right] \sqrt{1 - \kappa}   \nonumber \\
+ & 15 \kappa^5 \ln \frac{1 + \sqrt{1 - \kappa} }{1 + \sqrt{1 + \kappa}
} \ .
\end{align}
The above expressions have been obtained from (\ref{g1_def}) and
(\ref{g3_def}) without using any approximations. However, it should be
kept in mind that (\ref{g1_def}) is valid for $-1 \leq \kappa$, while
(\ref{g3_def}) is valid for $-1 \leq \kappa \leq 1$. Since
$\kappa=\hbar\omega /E_F$, we are interested only in the region
$\kappa>0$. The functions $g_1(\kappa)$ and $g_3(\kappa)$ in the
interval $0\leq \kappa \leq 1$ are shown in Fig.~\ref{fig:g1_g3}.

\end{document}